\documentclass[twocolumn]{aastex61}
\usepackage{amsmath}

\newcommand\aastex{AAS\TeX}

\received{}
\revised{}
\accepted{23-05-2018}
\submitjournal{ApJ}

\shorttitle{\aastex\ Forecasting Solar Eruptions via Machine Learning Methods}
\shortauthors{Inceoglu et al.}

\begin{document}

\title{Using Machine Learning Methods to Forecast If Solar Flares Will Be Associated with CMEs and SEPs}

\correspondingauthor{Fadil Inceoglu}
\email{fadil@eng.au.dk}

\author[0000-0003-4726-3994]{Fadil Inceoglu}
\affiliation{Department of Engineering, Aarhus University, Finlandsgade 22, DK-8200 Aarhus N, Denmark}
\affiliation{Department of Geoscience, Aarhus University, H{\o}egh-Guldbergs Gade 2, DK-8000 Aarhus C, Denmark}
\affiliation{Stellar Astrophysics Centre, Department of Physics and Astronomy, Aarhus University, Ny Munkegade 120, DK-8000 Aarhus C, Denmark}

\author{Jacob H. Jeppesen}
\affiliation{Department of Engineering, Aarhus University, Finlandsgade 22, DK-8200 Aarhus N, Denmark}

\author{Peter Kongstad}
\affiliation{Department of Geoscience, Aarhus University, H{\o}egh-Guldbergs Gade 2, DK-8000 Aarhus C, Denmark}

\author{N\'{e}stor J. Hern\'{a}ndez Marcano}
\affiliation{Department of Engineering, Aarhus University, Finlandsgade 22, DK-8200 Aarhus N, Denmark}

\author{Rune H. Jacobsen}
\affiliation{Department of Engineering, Aarhus University, Finlandsgade 22, DK-8200 Aarhus N, Denmark}

\author{Christoffer Karoff}
\affiliation{Department of Geoscience, Aarhus University, H{\o}egh-Guldbergs Gade 2, DK-8000 Aarhus C, Denmark}
\affiliation{Stellar Astrophysics Centre, Department of Physics and Astronomy, Aarhus University, Ny Munkegade 120, DK-8000 Aarhus C, Denmark}



\begin{abstract}

Among the eruptive activity phenomena observed on the Sun, the most technology threatening ones are flares with associated coronal mass ejections (CMEs) and solar energetic particles (SEPs). Flares with associated CMEs and SEPs are produced by magnetohydrodynamical processes in magnetically active regions (ARs) on the Sun. However, these ARs do not only produce flares with associated CMEs and SEPs, they also lead to flares and CMEs, which are not associated with any other event. In an attempt to distinguish flares with associated CMEs and SEPs from flares and CMEs, which are unassociated with any other event, we investigate the performances of two machine learning algorithms. To achieve this objective, we employ support vector machines (SVMs) and multilayer perceptrons (MLPs) using data from the Space Weather Database of Notification, Knowledge, Information (DONKI) of NASA Space Weather Center, {\it the Geostationary Operational Environmental Satellite} ({\it GOES}), and the Space-Weather Heliospheric and Magnetic Imager Active Region Patches (SHARPs). We show that True Skill Statistics (TSS) and Heidke Skill Scores (HSS) calculated for SVMs are slightly better than those from the MLPs. We also show that the forecasting time frame of 96 hours provides the best results in predicting if a flare will be associated with CMEs and SEPs (TSS=0.92$\pm$0.09 and HSS=0.92$\pm$0.08). Additionally, we obtain the maximum TSS and HSS values of 0.91$\pm$0.06 for predicting that a flare will not be associated with CMEs and SEPs for the 36 hour forecast window, while the 108 hour forecast window give the maximum TSS and HSS values for predicting CMEs will not be accompanying any events (TSS=HSS=0.98$\pm$0.02).

\end{abstract}

\keywords{Sun: magnetic activity, surface magnetic field, solar flare, coronal mass ejection, machine learning, SVM, MLP}



\section{Introduction} \label{sec:intro}

The Sun potentially endangers modern civilisation through large solar eruptions. Predicting and monitoring the large solar flares, coronal mass ejections (CMEs) and solar energetic particles (SEPs), which introduce enormous amounts of particles and energy to Earth's vicinity and into its atmosphere, is therefore of crucial importance.

The strongest observed flare and accompanying CME was the Carrington event that occurred in 1859, which was about twice as big as the strongest events observed during the space era \citep{1859MNRAS..20...13C,2013JSWSC...3A..31C}. Since the 1950s, a series of the most powerful flares and their accompanying CMEs have occurred between mid-October to early November 2003 peaking around 28 and 29 October (the so-called Halloween solar storms), which caused radio blackouts on Earth, and problems in different instruments of the satellites, such as star trackers. These flares even killed the power supply of the Japanese Earth-resource satellite, the MIDORI-II (ADEOS-2), and left it inoperative. The effects of the Halloween solar storms extended beyond the Earth to Mars and caused the Mars Odyssey spacecraft to go into deep safe-mode \citep{2004EOSTr..85..105L}. 

The CMEs originate from large coronal loop structures, which contain plasma and magnetic fields, expanding from the Sun into the interplanetary medium. They occur on a quasi-regular basis and are the largest-scale eruptive phenomena in our solar system \citep{2011LRSP....8....1C,2012LRSP....9....3W,2017LRSP...14....5K}. Flares, on the other hand, are eruptive events that occur in the solar atmosphere with energies ranging from 10$^{28}$ to 10$^{32}$ erg \citep{2011LRSP....8....6S}. Based on their peak fluxes in the 1 to 8 Angstrom range X-rays near Earth, as measured by XRS instrument on the {\it Geostationary Operational Environmental Satellite} ({\it GOES}), flares are classified as A, B, C, M, and X. Each flare class is also divided into subclasses linearly scaled from 1 to 9, such as M4 or X9 \citep{2010hssr.book.....S}. Strong flares and CMEs occasionally cause SEP events, where protons, electrons, and heavier nuclei are accelerated to the energies ranging between a few tens of keVs to GeVs \citep{2013SSRv..175...53R}. Strong SEPs cause nuclear cascades in the upper atmosphere of the Earth \citep{2013SoPh..285..233R}. Flares are often associated with CMEs as they are thought to have a common magnetically-driven mechanism and not one causing the other \citep{2012LRSP....9....3W}. However, there are CMEs and flares that are not associated with one another \citep{2011LRSP....8....1C,2012LRSP....9....3W}. Observations show that the speeds and energies of the CMEs are higher when they are accompanied by bright flares in comparison to those not accompanying with flares \citep{2011LRSP....8....1C,2012LRSP....9....3W}.

Magnetically strong regions on the photosphere, the so-called active regions (ARs), are often the source regions to flares and CMEs \citep{2011LRSP....8....1C,2015LRSP...12....1V}. Strong ARs are generally large and evolve rapidly with lifetimes varying from days to months \citep{2015LRSP...12....1V}, and they exhibit complex magnetic geometry \citep{2008LRSP....5....1B}. Vector magnetograms, which measure the line-of-sight (LoS) magnetic field separately from the image-plane, allow us to calculate the physical indices of the ARs, such as magnetic helicity, magnetic shear angles, proxies for free energy and magnetic fluxes of the ARs, and polarity-separation lines \citep{2003ApJ...595.1277L,2007ApJ...655L.117S,2012ApJ...750...24M}.

Previous studies on predicting solar eruptive phenomena used photospheric magnetic field data to calculate the physical indices of ARs and they link these physical indices to the occurrences of flares and CMEs \citep{2003ApJ...595.1277L, 2007ApJ...655L.117S, 2012ApJ...750...24M}. Recently, machine learning (ML) methods, such as support vector machines (SVMs) and neural networks (NNs), have been used in predicting flares, CMEs, and SEPs \citep{2009SoPh..255...91Y,2010RAA....10..785Y,2013SoPh..283..157A,2015ApJ...812...51B,2015ApJ...798..135B,2016ApJ...821..127B,2018SoPh..293...28F}. 

\citet{2009SoPh..255...91Y} used LoS magnetogram data from the Michelson Doppler Imager (MDI) on {\it the Solar and Heliospheric Observatory} ({\it SOHO}) to predict flare occurrences within a forecast window of 48 hours. To achieve this objective, the authors used data from ARs, which generated at least one $\geq$C1 class flare. They divided this data into two subclasses as flaring and non-flaring regions based on their total importance threshold value \citep[Equation 1 in][]{2009SoPh..255...91Y}, and used it in a learning vector quantisation neural networks algorithm. \citet{2010RAA....10..785Y}, on the other hand, used the same data to predict flare classes A, B, C, M, and X via SVMs.

To predict $>$C1 class flares 24 and 48 hours prior to their occurrences, \citet{2013SoPh..283..157A} used magnetic feature data of flaring and non-flaring regions from the Helioseismic and Magnetic Imager (HMI) on the {\it Solar Dynamics Observatory} ({\it SDO}) in a cascade correlation neural network algorithm. The authors defined an AR as non-flaring if it does not produce a flare within 24 hours for their 24 hour prediction window, while for the 48 hour forecast window they defined an AR as non-flaring if it does not produce a flare within $\pm$48 hours after the sampling time. \citet{2015ApJ...812...51B}, on the other hand, used the {\it SOHO}/MDI data of flaring and non-flaring ARs in a SVM regressor.

To predict occurrences of flares larger than M1 class, \citet{2015ApJ...798..135B} used {\it SDO}/HMI's definitive flaring and non-flaring AR data, which are defined similar to \citet{2013SoPh..283..157A}, in SVMs at time delays 48 and 24 hours, respectively. In addition, \citet{2016ApJ...821..127B} used the {\it SDO}/HMI's definitive AR data that produce flares and flares with associated CMEs in SVMs to predict whether a flare will be associated with CMEs within a 24 hour forecast window.

\citet{2018SoPh..293...28F}, on the other hand, calculated physical features of flaring and non-flaring ARs, which they identified on the {\it SDO}/HMI's near-real-time vector magnetogram data. The flaring and non-flaring ARs are defined based on whether they produce a flare within a 24 hour forecasting window or not. The authors used this data in SVMs, multilayer perceptrons (MLPs), which is based on NNs, and also decision tree algorithms to predict occurrences of $>$M1-class and $>$C1-class flares.

In this study, we aim to distinguish flares with associated CMEs and SEPs from flares and CMEs without any accompanying events. To predict which event will be produced from an AR, all of which lead to solar eruptions, we investigate the performances of two ML algorithms, SVMs and MLPs, based on the vector magnetic field data observed with the HMI on the {\it SDO} \citep{2012SoPh..275..229S}. The results will also highlight the discriminative potential of the Space-Weather Heliospheric and Magnetic Imager Active Region Patches (SHARPs) data among the three classes. Section~\ref{sec:data} describes the data used in this study, while the performed analyses, including brief explanations of the ML methods, are presented in Section~\ref{sec:analyses}. Results from the analyses are shown in Section~\ref{sec:results} and discussion and conclusions are given in Section~\ref{sec:dis_conc}.

\section{The {\it SDO}/HMI Data}
\label{sec:data}

To predict which solar eruptive phenomena will be generated from an AR, which is known to have generated only flares, flares with associated CMEs and SEPs, or only CMEs, via SVMs and MLPs, we use data from the Space Weather Database Of Notification, Knowledge, Information (DONKI)\footnote{\url{http://kauai.ccmc.gsfc.nasa.gov/DONKI/}} of NASA Space Weather Research Center, for a period spanning from 01 January 2010 to 31 January 2018. DONKI contains flare data with their classes, source AR numbers, and their start, peak, and end times. This database also provides whether a flare is accompanied with CMEs and/or SEPs. Similar to the flare data, DONKI also provides CME data with their speeds, types, and start times as well as whether they are associated with any flares and/or SEPs. For some events however, DONKI does not provide an AR number although an M or X class flare with an accompanying CME is listed. For example, the M3.0 class flare with an accompanying CME and SEP that occurred on 06 March 2015. To fill the unregistered AR numbers in DONKI for these events and also to double check the peak times and the AR numbers of flares, we use flare data from {\it GOES} via SunPy Python package v0.8.2 \citep{2015CS&D....8a4009S}.

\begin{figure}
\begin{center}
{\includegraphics[width=3in]{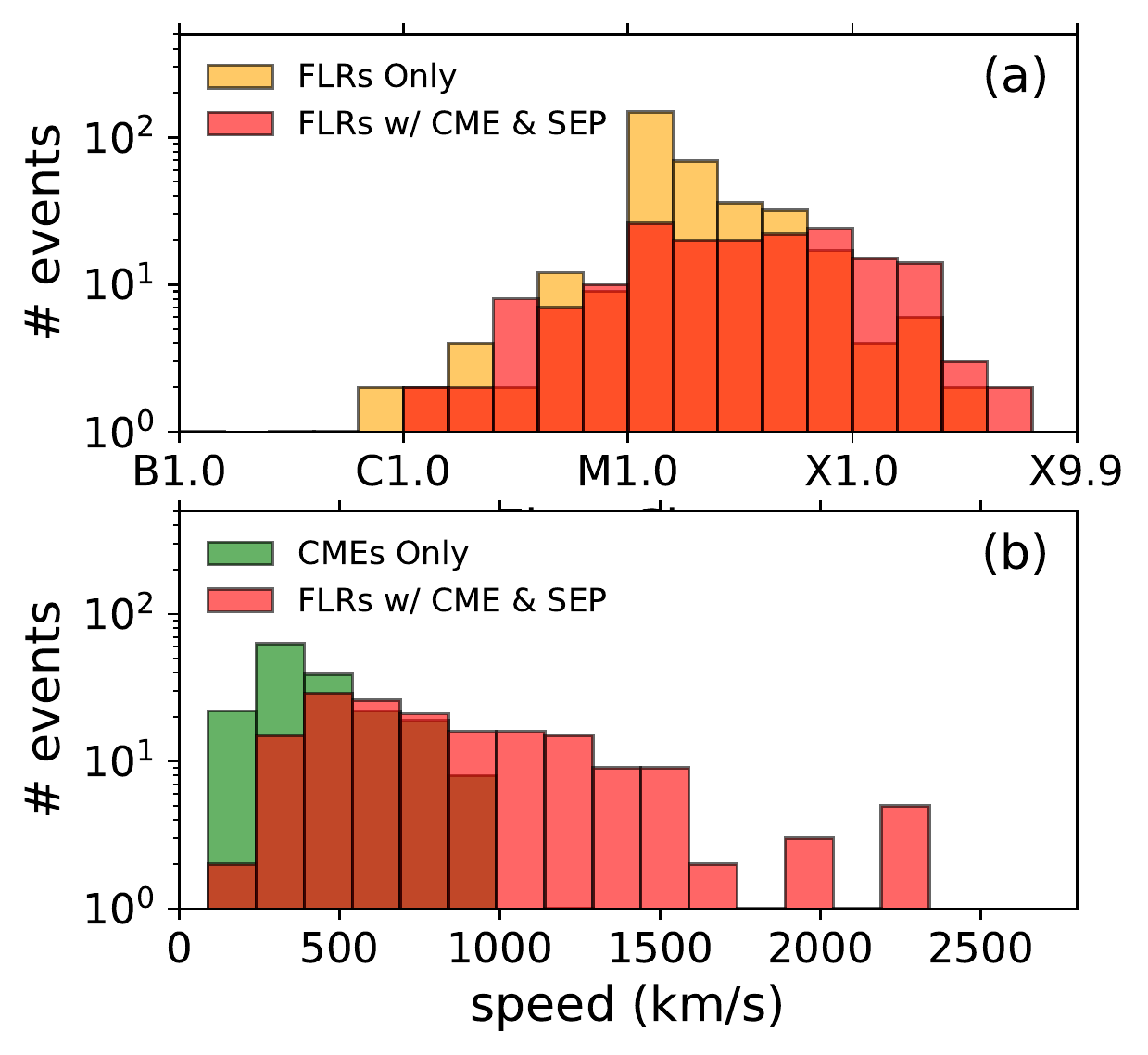}}
\caption{The top panel shows the distributions of the flare classes of flares (orange) and flares with CMEs and SEPs (red), while bottom panel shows the distributions of the CME speeds for CME only events (green) and flares with CMEs and SEPs (red). Note that the y-axes in all panels are in logarithmic scale.}
\label{fig:data}
\end{center}
\end{figure}

Based on the DONKI and the {\it GOES} databases we have 347 flares that are not associated with CMEs and SEPs, and 179 flares associated with CMEs and SEPs (Figure~\ref{fig:data}a). Further, there are 376 CMEs in total, 174 of which are not associated with any other event (Figure~\ref{fig:data}b).

Following the verification of the flare data from the DONKI and the {\it GOES} databases, we use publicly available SHARPs data from the Joint Science Operations Center (JSOC)\footnote{\url{http://jsoc.stanford.edu}}, spanning from 01 January 2010 to 31 January 2018. We select the SHARPs data based on four criteria following \citet{2016ApJ...821..127B}, so the SHARPs data has to be; (i) disambiguated, (ii) taken while the {\it SDO}'s orbital velocity $<3500 m\ s^{-1}$, (iii) of a high-quality, meaning the data with reliable Stokes vectors (observables during good conditions), and (iv) within $\pm70^{\circ}$ longitudinal band during {\it GOES} peak time, since beyond this band the signal-to-noise ratio in the vector magnetic field data decreases significantly. The SHARPs data contain vector magnetic field measurements of the ARs and 18 keywords, which are listed in Table~\ref{tab:data} together with their definitions. These keywords parametrise the measured physical quantities as well as proxies of physical quantities \citep[for details, see][]{2014SoPh..289.3549B}. 

\begin{table}[h!]
\centering
\caption{Keywords, definitions, and formulations of the 18 physical features of ARs. The {\it Keyword} column indicates the name of the FITS header keyword in the SHARP data series.} 
\label{tab:data}
\begin{tabular}{ll}
\hline
\hline
\multicolumn{1}{l}{Keyword} & \multicolumn{1}{c}{Definition}  \\
\hline
ABSNJZH & Absolute value of the net current helicity       \\
R\_VALUE &  Sum of flux near polarity inversion line     \\
AREA\_ACR & Area of strong field pixels \\
			& in the active region     \\
MEANSHR & Mean shear angle     \\
TOTPOT & Total photospheric magnetic \\
		& free energy density  \\
SAVNCPP & Sum of the modulus of the net current \\
		& per polarity      \\
TOTUSJZ & Total unsigned vertical current     \\
MEANJZD &  Mean vertical current density   \\
MEANGBZ & Mean gradient of vertical field     \\
MEANGAM & Mean angle of field from radial    \\
MEANALP & Mean characteristic twist parameter, $\alpha$    \\
MEANGBH &  Mean gradient of horizontal field     \\
TOTUSJH &  Total unsigned current helicity       \\
SHRGT45 &  Fraction of Area with Shear $> 45^{\circ}$    \\
MEANPOT & Mean photospheric magnetic free energy    \\
MEANJZH &  Mean current helicity ($B_z$ contribution)     \\
MEANGBT & Mean gradient of total field     \\
USFLUX &  Total unsigned flux     \\
\hline
\end{tabular}
\end{table}

The 18 physical features of the ARs (Table~\ref{tab:data}) for the three subsets are then compiled at a time before the starting times of the events. This time gap is defined as the time-delay ($\Delta t$), which ranges from 12 to 120 hours with 12 hour intervals. This means that we use conditions $\Delta t$ hours before the event occurs to predict whether this event will be a flare without associated events, flare with associated CMEs and SEPs or CMEs without associated events. For each time delay iteration, the data from the DONKI and {\it GOES} databases goes through the same data selection criteria and this causes the sample size of each three class to change (Table~\ref{tab:datasize}). This situation is directly related to the temporal evolution and motion of the ARs on the solar disk as well as availability of the SHARPs data for a given time time delay.

\begin{table}[h!]
\centering
\caption{Number of flares, flares with CMEs and SEPs, and CMEs for each time delay ranging from 12 to 120 hours.} 
\label{tab:datasize}
\begin{tabular}{lccc}
\hline
\hline
\multicolumn{1}{l}{time delay} & \multicolumn{1}{c}{flares} & \multicolumn{1}{c}{flares w/ CMEs \& SEPs} & \multicolumn{1}{c}{CMEs}  \\
\hline
				12 h 			& 228			& 103								& 97       \\
				24 h 			& 237			& 100								& 94       \\
				36 h 			& 239			& 105								& 102       \\
				48 h 			& 236			& 100								& 99       \\
				60 h 			& 226			& 101								& 93       \\
				72 h 			& 215			& 95									& 88       \\
				84 h 			& 205			& 91									& 87       \\
				96 h 			& 196			& 90									& 87       \\
				108 h 		& 182			& 80									& 79       \\
				120 h 		& 177			& 76									& 76       \\												

\hline
\end{tabular}
\end{table}

The flares that occurred during the study period cluster in M-class flares (Figure~\ref{fig:data}a). The flares unassociated with any other events centre around M1.0 class, whereas flares associated with CMEs and SEPs cluster around M5.0-class and extend up to X9.9 (Figure~\ref{fig:data}a).

The distribution of the CME speeds ranges between 90 to 2650 km\ s$^{-1}$. The speeds of the CMEs that are unassociated with an event centre around 500 km\ s$^{-1}$, while the speeds of the CMEs associated with flares and SEPs centre around 1000 km\ s$^{-1}$ and reach up to 2650 km\ s$^{-1}$ (Figure~\ref{fig:data}b).

The distribution of the flare classes and the CME speeds indicate that we can separate our data into three subsets as (i) flares only, (ii) flares with associated CMEs and SEPs, and (iii) CMEs only. These three subsets represent the classes for our multi-class classification problem, which we aim to sort out using the SVM and MLP algorithms. It must be noted that the size of the underlying data from the {\it SDO}/HMI is smaller than those used in previous studies based on the {\it SOHO}/MDI, mission duration of which has reached $\sim$22 years. Additionally, the current solar cycle 24 is quiet in nature and does not generate many eruptions compared to previous stronger cycles.

\section{Analyses}
\label{sec:analyses}

Before using the 18 physical parameters of the ARs in further analyses for each $\Delta t$ hours delay, we standardise all the data according to their median and standard deviation values. The reason for this approach is to make sure that the data with small sample size is well represented, the distribution of which might sometimes be left or right skewed. For larger sample sizes, the median and the mean values overlap.

\subsection{Machine Learning Algorithms}

To investigate which ML method provides better predictions of our three classes, we use two of the most popular machine learning algorithms; (i) SVMs \citep{CortesSVMOrig1995}, and (ii) MLPs \citep{HornikWhiteMLP89} provided by the {\it scikit-learn} software package v0.19.1 \citep{Pedregosa2011}.

\subsubsection{Support Vector Machines}
\label{sec:SVMs}

SVMs are initially designed to solve binary ($l=2$) classification problems and employing them to multi-class ($l>2$) classification problems requires different approaches, where they are generally fragmented into series of different binary classification problems \citep{HsuandLin2002}. In this study, we use the one-versus-rest approach \citep{Vapnik1998book}, which creates $k$ separate binary classifiers for $l$ number of classes. The $m$-th binary SVM classifier is then trained using the data from the $m$-th class as positive ($+$1), whereas the remaining $l-1$ number of classes are regarded as negative ($-$1) examples \citep{HsuandLin2002}. The SVM then classifies the data by placing a separating hyperplane with the maximum distance between the classes of the data. 

Let us consider a multi-class classification reduced to a binary class problem via one-versus-rest approach for an $m$-th class. For a vector of $P$ predictor consisting of training data at observation $i$ is given as a pair $(x_i, y_i)$, where $i=1,..,n$, and $x_i  \in \mathbb{R}^P$ and its class $y_i \in \{+1, -1\}$, then the SVM solves the optimisation problem as follows \citep{HsuandLin2002},

\begin{eqnarray}
\label{eq:SVM1}
&\min&_{w, b, \zeta} \ \frac{1}{2}\omega^T \omega +C \sum\limits_{i=1}^n\zeta_i,  \\
\text{subject to} &y_i&(\omega^T\phi(x_i)+b) \ge 1-\zeta_i, \nonumber \\  \nonumber
&\zeta_i& > 0, \ i=1,...,n, \\  \nonumber
\end{eqnarray}

\noindent where $\omega$ is the weight vector, while $\phi(x_i)$ is an unknown function included in a known {\it kernel} function that maps the training vectors $x_i$ into a higher dimensional space. $C>0$ is the regularisation parameter, which compromises misclassification of training examples to make the decision surface simpler. Lower $C$ values make the decision surface smooth, while higher values enable the algorithm to minimise the errors on the training examples and maximise the separation margins \citep{CortesSVMOrig1995}. The kernel function is defined as the inner product of the data with itself for different pairs of observations $i$ and $j$, $K(x_i,x_j)=\phi(x_i)^T\phi(x_j)$ \citep{2018SoPh..293...28F}. In our study, we use a Radial Basis Function (RBF) kernel that is defined as,

\begin{equation}
K(x_i,x_j)=\exp(-\gamma||x_i-x_j||^2),
\label{eq:kernelSVM}
\end{equation}

\noindent where $\gamma >0$ defines how much influence a single training example have on the classification. If $\gamma$ is a large value, then the other examples need to be closer to be influenced, whereas smaller values can make the model too constrained, which causes the model to not capture the complexity of the underlying data. A more detailed information on how the solve Equations~\ref{eq:SVM1} and~\ref{eq:kernelSVM} can be found elsewhere \citep{CortesSVMOrig1995,Vapnik1998book}. 

\subsubsection{Multilayer Perceptrons}

The MLP is a feed-forward NN that classifies multi-dimensional data into $l$ different classes. The multi-class classification problem in MLPs can be regarded as a multinomial logistic regression, where the output of the NN is the posterior probability that the input data belongs to a particular class. The estimated posterior probability distribution of a categorical random variable depends not only on a data point from a random feature, but also on the weights of the neurons, which are the basic processing units \citep{MacKay2005Book}. As a feed-forward network, the MLP performs nonlinear parametrised mapping from an input $\boldsymbol{I}$ to an output that is a continuous function of the input and the weights $\boldsymbol{O=(I;\omega,A)}$, where $\boldsymbol{\omega}$ is the weight and bias parameter vector, and $\boldsymbol{A}$ is the architecture of the network defining the number of nodes in every layer \citep{2018SoPh..293...28F}.

In this study, our MLPs have an input layer with $m=18$ inputs $I_m$ and bias $\theta^{(1)}$, a single hidden layer with $j$ hidden nodes $H_{j}$ and a bias $\theta^{(2)}$, and an output layer $O_i$ that only post-processes a data point to give an estimate of the posterior probability. The architecture is given below following \citet{MacKay2005Book} and \citet{2018SoPh..293...28F};

\begin{eqnarray}
\label{eq:mlp}
&&\alpha_j^{(1)}=\sum\limits_{m} \omega_{jm}^{(1)} I_m+\theta_j^{(1)}; \ H_{j}=f(\alpha_j^{(1)}), \\ \nonumber
&&\alpha_i^{(2)}=\sum\limits_{j} \omega_{ij}^{(2)} H_{j}+\theta_i^{(2)}; \ O_{i}=g(\alpha_i^{(2)}), \nonumber
\end{eqnarray}

\noindent where $f(\alpha)=\frac{1}{(1+exp(-\alpha))}$ is the logistic activation function, which defines the response of a neuron in the hidden layer to a stimulus obtained from its input \citep{Hinton1989}. The index $m$ used for the inputs, $j$ and $i$ run over the hidden units and outputs, respectively. The weights $\omega$ and biases $\theta$ define the parameter vector $\boldsymbol{\omega}$ \citep{MacKay2005Book}.

The function $g(\alpha_i)$ coupled to the output layer $O_i$ is called the softmax activation function because our MLP is designed to solve a multi-class classification problem, which is given as \citep{MacKay2005Book};

\begin{equation}
\label{eq:softmax}
g(\alpha_i)=\frac{\exp(\alpha_i)}{\sum_{l=1}^k \exp(\alpha_l)},
\end{equation}

\noindent where $\alpha_i$ represents the $i$-th element of the input to softmax corresponding to class $l$, and $k$ is the number of classes. 

The MLPs are trained using a data $D = \{\boldsymbol{I}^{(n)}; t^{(n)}\}$ by adjusting $\boldsymbol{\omega}$ to minimise the negative log-likelihood function given as \citep{MacKay2005Book};

\begin{equation}
\label{eq:log_loss}
G(\boldsymbol{\omega}) = -\sum_{n}\sum_{i} t_i^{(n)}\ln\boldsymbol{O_i}(\boldsymbol{I}^{(n)};\boldsymbol{\omega}),
\end{equation}

\noindent where $\boldsymbol{I}^{(n)}$ is the predictor matrix, $t_i^{(n)}$ is the target vector, and $n=1, \ldots, N$ is the observations. 

Minimising the log-loss function is equivalent to obtaining the maximum likelihood estimator of the weights and biases. To find the output and hidden layer weights and biases, we use a L-BFGS unconstrained optimisation algorithm. 

\subsection{Training and tuning of the algorithms}

The size of the underlying dataset is limited because of the quiet solar cycle 24 and the mission duration of the {\it SDO}/HMI. This situation leads to difficulties in reaching a sufficient number of events to separate the data into training and testing subsets. To overcome this difficulty, we employ the stratified $k$-fold cross-validation (CV) method, following \citet{2016ApJ...821..127B}, where we change $k$-fold values from 3 to 15. This method divides the dataset into $k$ subsets, where one subset is used for test and the remaining are used for training. This process is then repeated $k$ times, such that each subset is used for testing exactly once, before the results are averaged. 

Limitations of using this method do exist, however, as it assumes that the dataset to follow the same distribution over time. In our case, the number of events varies in-phase with the 11-year sunspot cycle, which is expected considering the formation of the ARs on the solar photosphere by emergence of deep-seated toroidal magnetic flux ropes, which are generated by the shearing of the poloidal magnetic fields by solar differential rotation, through the photosphere to the corona \citep{2010LRSP....7....3C,2015LRSP...12....1V}. However, significant flares can occur at all phases of the sunspot cycle, as three of the last four solar cycles showed X-class flares around their minima \citep{2015LRSP...12....4H}. The date of the event is not included in the classifier, thus this bias should be accounted for. This is done by shuffling the data before the $k$-fold cross-validation is carried out. 

In addition to employing the $k$-fold cross-validation, we tune the hyper-parameters of the SVMs and MLPs in each $k$-fold. To achieve this, we use the grid search algorithm provided by the {\it scikit-learn} software package \citep{Pedregosa2011}. This algorithm performs an exhaustive search over a predefined set of hyper-parameter values and the best combination of them is then retained. For example, two hyper-parameters with lists of predefined values would result in a 2D grid, where each element is tested and the pair of values that gives the best result is saved. While training the SVM, the regularisation parameter, $C$, vary between $10^{-3}$ and $10^{7}$, while the radial basis function coefficient, $\gamma$, vary from $10^{-6}$ to $10^{4}$ in 11 equidistant values in log-space. For the MLP, we have an input, a hidden, and an output layer, where the size of the hidden layer is tested with $[18, 36, 54, 72, 90, 108]$ neurons. Additionally, we vary both the regularisation and tolerance parameters between 10$^{-7}$ and 10$^{-3}$ in 5 equidistant values in log-space.

\subsection{Comparison of the Algorithms}
\label{sec:CompAlgAnalysis}

Most of the metrics that are used to characterise and quantify the predictive power of classifiers are calculated directly from the confusion matrices, which are built based on the results from the raw classifier outputs. A confusion matrix for $l$ number of classes shows how $n$ number of instances are distributed over predicted $P_{i}$ and observed $O_{i}$ classes, where $1\le i \le l$.

\begin{table}[h!]
\centering
\caption{A confusion matrix. The term $n_{ij}$ denote the number of instances predicted in class $i$ by the classifier ($P_{i}$), where it is observed in class $j$ ($O_{j}$), and $1 \le i, j \le l$.}
\label{tab:conf_matrix}
\begin{tabular}{c|ccc}

\hline
									& $O_{1}$ & $\cdots$ & $O_{l}$  \\
\hline
				$P_{1}$ 				& $n_{11}$	&$\cdots$			& $n_{1l}$	     \\
				$\vdots$	 			& $\vdots$	&$\ddots$			& $\vdots$	   \\
				$P_{l}$ 				& $n_{l1}$		&$\cdots$			& $n_{ll}$	      \\

\hline
\end{tabular}
\end{table}

The diagonal terms in a confusion matrix ($i=j$) show correctly predicted instances, while the off-diagonal terms ($i \neq j$) represents incorrectly predicted classes (Table~\ref{tab:conf_matrix}). 

Let us now consider only one class $i$. The confusion matrix obtained from a classifier gives four types of instances, which are true positives (TP), false positives (FP), false negatives (FN), and true negatives (TN). TP and FP refer to the events that are predicted and observed, and that are predicted but not observed, respectively. FN and TN, on the other hand, represent the events that are not predicted but observed, and that are not predicted and not observed, respectively. For the class $i$, these metrics are calculated as $TP=n_{i,i}$, $FP=n_{i,+}-n_{ii}$, $FN=n_{+,i}-n_{i,i}$ and $TN=n-TP-FP-FN$, where $n_{i,+}$, and $n_{+,i}$ denote the sums of the confusion matrix elements over row $i$ and column $i$, respectively \citep{Labatut2011}.

To evaluate the predictive power of the algorithms, (i) SVMs for different $k$-folds, (ii) MLPs for different $k$-folds, and (iii) SVMs versus MLPs for the same $k$-fold, we use the True Skill Statistics (TSS) \citep{HanssenKuipersTSS} and the Heidke Skill Score (HSS) \citep{Heidke1926}.

The TSS compares the probability of detection (POD), to the probability of false detection (POFD) and is calculated based on the confusion matrices obtained from ML-algorithms. The TSS is defined as,

\begin{eqnarray} 
\label{eq:TSS}
TSS&=& POD-POFD, \\ \nonumber
&=& \frac{TP}{TP+FN}-\frac{FP}{FP+TN},
\end{eqnarray}

The TSS varies between $-1$ and $+1$, and the value of 0 means that the algorithm is incompetent. High positive values indicate that the algorithm performs well, while negative values show a contradictory behaviour, suggesting that the positive and negative classes are mixed around and therefore giving a reversed score \citep{2018SoPh..293...28F}. 

The HSS, on the other hand, is a method of measuring the fractional improvement of the forecast over the random forecast and it is defined as,

\begin{eqnarray}
\label{eq:HSS}
&&HSS=\nonumber \\ 
&&\frac{2(TP \times TN-FP \times FN)}{(TP+FN)(FN+TN)+(TP+FP)(FP+TN)}, \nonumber \\
\end{eqnarray}

The HSS ranges from $- \infty$ to 1. A negative value in the HSS method, means that the random forecast is better. A value of 0 means that the model has no skill over the random forecast, while positive values indicate the optimum forecasting method \citep{2018SoPh..293...28F}.

Both the TSS and HSS values are calculated based on the confusion matrices obtained from the SVMs and MLPs for each $k$-fold CV iteration, ranging from 3 to 15. This method, in turn, will provide $n$ TSS and $n$ HSS values for $n$-folds cross-validation, from which averages and standard deviations of TSS and HSS can be calculated; for example, for a 10-fold cross-validation, we will calculate 10 TSS and 10 HSS values. The results presented in the next section are based on the averaged TSS and HSS values. 

\section{Results}
\label{sec:results}

\subsection{Machine Learning Algorithms}

To predict which solar eruptive phenomena will be generated from an AR, which is known to have generated flares alone, flares with associated CMEs and SEPs, or CMEs alone, at $\Delta t$ hour before they occur, we train SVMs and MLPs separately. Previously, it was suggested that using 12 to 18 features will not result in overfitting and using a single best feature will only cause marginally better results \citep{2016ApJ...821..127B}, therefore we did not include any feature selection criterion for our algorithms.

\subsubsection{Support Vector Machines}
\label{sec:resultsSVM}

\begin{figure}[ht!]
\begin{center}
{\includegraphics[width=3.5in]{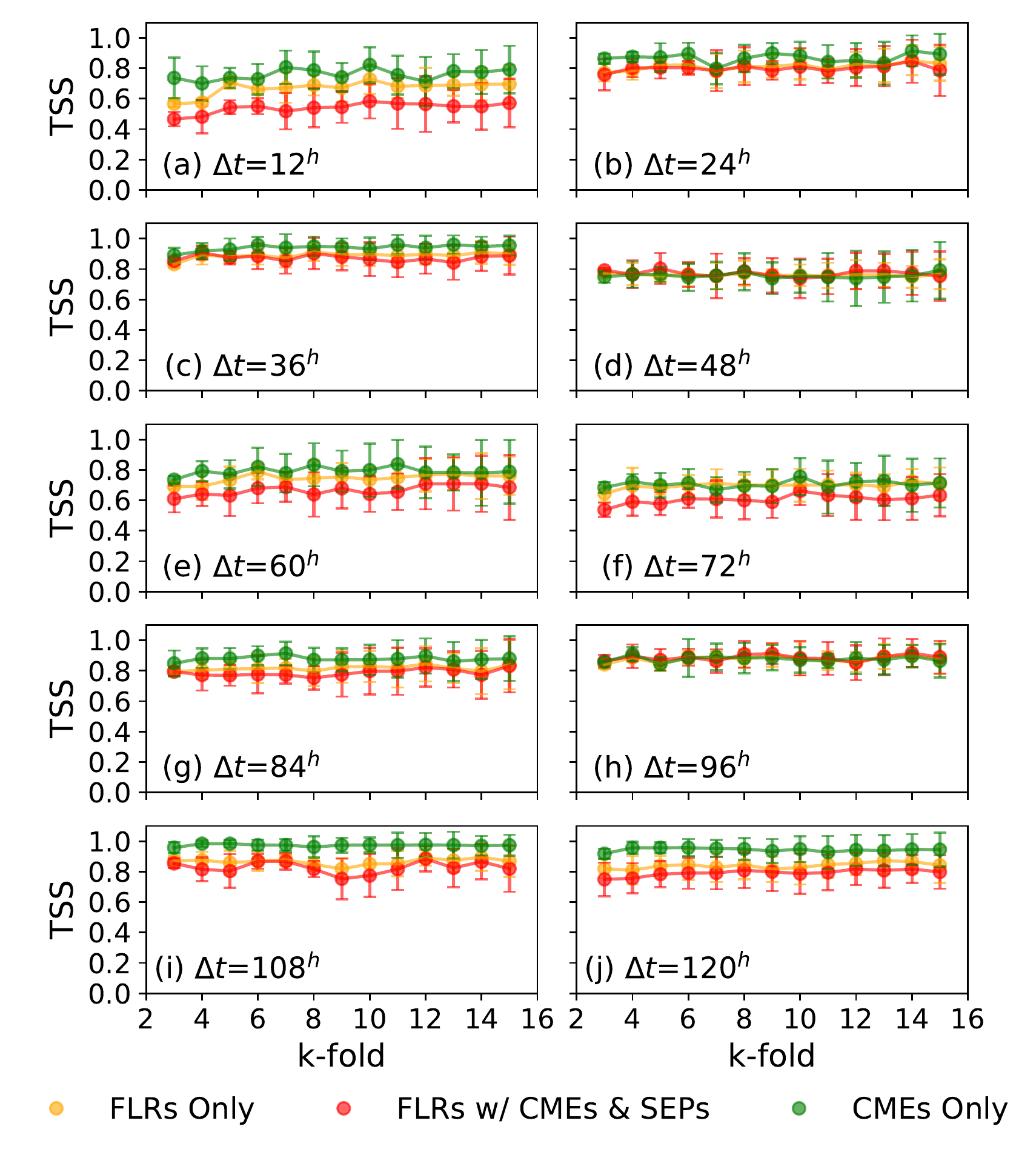}}
\caption{The TSS values calculated based on confusion matrices obtained from the SVMs as a function of $k$-folds CV and $\Delta t$ time delays for flares only (orange), flares with CMEs and SEPs (red), and CMEs only (green).} 
\label{fig:TSS_SVM}
\end{center}
\end{figure}

\begin{figure}[ht!]
\begin{center}
{\includegraphics[width=3.5in]{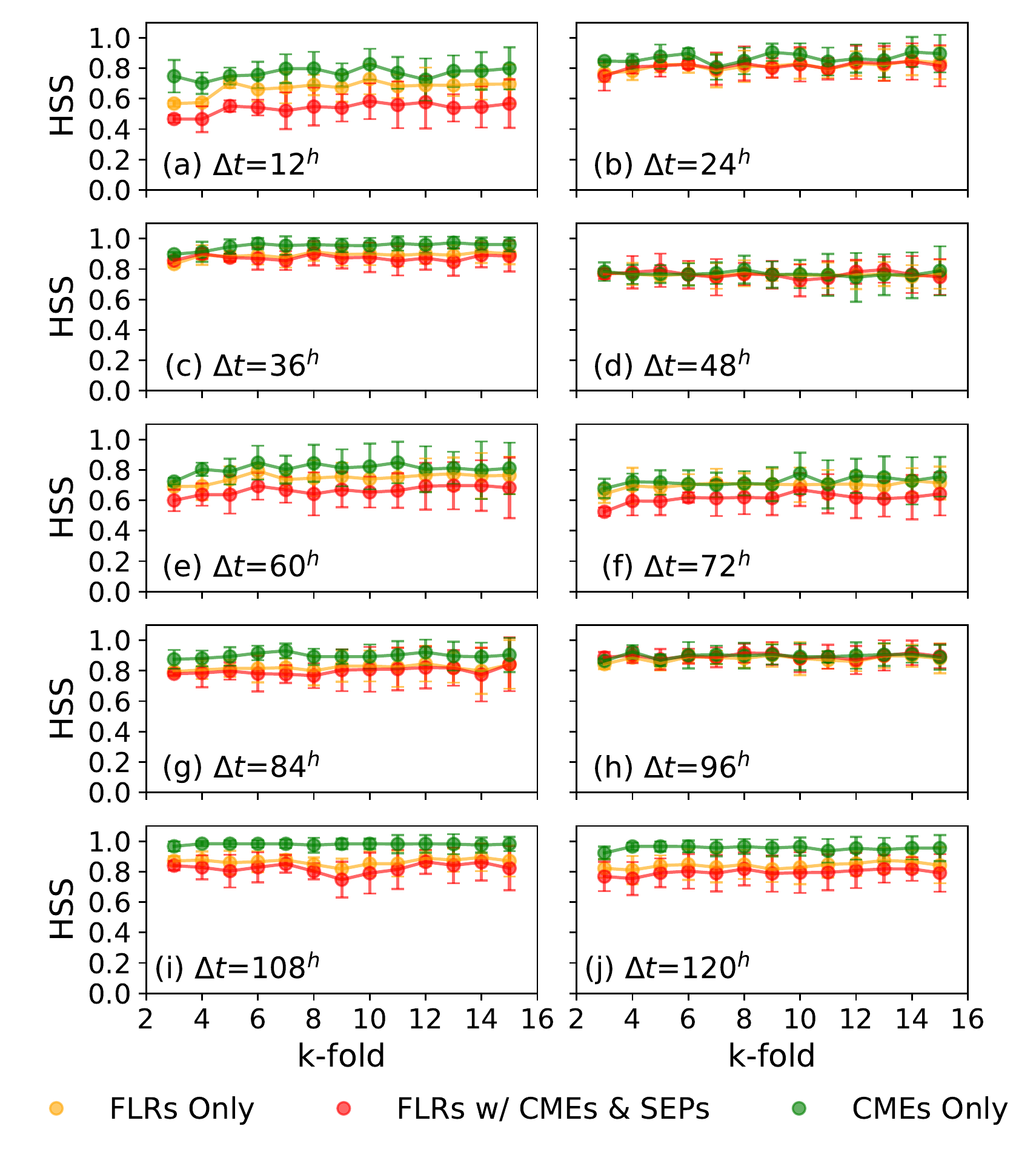}}
\caption{The HSS values calculated based on confusion matrices obtained from the SVMs as a function of $k$-folds CV and $\Delta t$ time delays for flares only (orange), flares with CMEs and SEPs (red), and CMEs only (green).} 
\label{fig:HSS_SVM}
\end{center}
\end{figure}

The TSS and HSS values are calculated based on the confusion matrices obtained from the SVMs for different $k$-fold CV. A sample for the confusion matrix for the 36 hour delay and $k=3$-fold CV show that 74 instances of the flares, 32 of the flares with CMEs and SEPs, and again 32 of the CMEs are correctly classified (Table~\ref{tab:conf_matrixSVM}). The results from TSS values suggest that we can predict if a flare will be unassociated with any other events 96 and 36 hours prior to their occurrences, where the maximum TSS level of 0.91$\pm$0.06 is found for the 36 hour delay. This value decreases down to $\sim$0.76 on average for the time delays between 48 and 72 hours and increases to around 0.85 at the 84 hour delay and reaches $\sim$0.88$\pm$0.08 level for the 108 and 120 hour delays. The maximum TSS values for the 24 and 12 hour delays are 0.85$\pm$0.10 and 0.73$\pm$0.10, respectively (Figure~\ref{fig:TSS_SVM} and Table~\ref{tab:MaxTSS_HSS_SVM}). A similar variation pattern can also be observed in the HSS values (Figure~\ref{fig:HSS_SVM} and Table~\ref{tab:MaxTSS_HSS_SVM}).

\begin{table}[h!]
\centering
\caption{A sample confusion matrix for one of the SVMs for the 36 hour delay, where $C=10^{4}$, $\gamma=10^{-3}$, and $k=3$-fold CV. Indices 1, 2, and 3 denote flare, flare with CMEs and SEPs, and CMEs classes, respectively.}
\label{tab:conf_matrixSVM}
\begin{tabular}{c|ccc}

\hline
									& $O_{1}$ 	& $O_{2}$ 		& $O_{3}$  	\\
\hline
					$P_{1}$ 			& 74			&3				& 1		     \\
					$P_{2}$	 		& 3			&32				& 1			   \\
					$P_{3}$ 			& 3			&0				& 32		      \\

\hline
\end{tabular}
\end{table}

\begin{table*}[ht!]
\centering
\caption{Maximum TSS and HSS values from the SVMs for three classes at different time delays. Note that the TSS and HSS values $\ge$\ 0.90 with standard deviation $\le$\ 10\% are marked in bold face.} 
\label{tab:MaxTSS_HSS_SVM}
\begin{tabular}{ccccccccccc}
\hline
\hline
\multicolumn{1}{c}{time delay} & \multicolumn{2}{c}{flares} & \multicolumn{2}{c}{flares w/ CMEs \& SEPs} & \multicolumn{2}{c}{CMEs}   \\

\multicolumn{1}{c}{$\Delta t$ (hrs)} & TSS$_{Max}$ &HSS$_{Max}$   & TSS$_{Max}$ &HSS$_{Max}$ & TSS$_{Max}$ &HSS$_{Max}$     \\
\hline

12 & 0.73$\pm$0.10 & 0.73$\pm$0.10 & 0.58$\pm$0.11 & 0.58$\pm$0.12 & 0.82$\pm$0.11 & 0.83$\pm$0.10  \\
24 & 0.85$\pm$0.10 & 0.85$\pm$0.10 & 0.85$\pm$0.14 & 0.84$\pm$0.12 & {\bf 0.91$\pm$0.10} & {\bf 0.91$\pm$0.10}  \\
36 & {\bf 0.91$\pm$0.06} & {\bf 0.91$\pm$0.06} & {\bf 0.91$\pm$0.05} & {\bf 0.90$\pm$0.08} & {\bf 0.96$\pm$0.06} & {\bf 0.97$\pm$0.04}  \\
48 & 0.77$\pm$0.08 & 0.77$\pm$0.08 & 0.80$\pm$0.10 & 0.80$\pm$0.09 & 0.79$\pm$0.19 & 0.80$\pm$0.09  \\
60 & 0.79$\pm$0.05 & 0.79$\pm$0.05 & 0.71$\pm$0.18 & 0.70$\pm$0.17 & 0.84$\pm$0.16 & 0.85$\pm$0.14  \\
72 & 0.72$\pm$0.08 & 0.73$\pm$0.08 & 0.66$\pm$0.09 & 0.67$\pm$0.11 & 0.76$\pm$0.12 & 0.78$\pm$0.14 \\
84 & 0.85$\pm$0.11 & 0.85$\pm$0.12 & 0.83$\pm$0.18 & 0.84$\pm$0.18 & {\bf 0.91$\pm$0.08} & {\bf 0.93$\pm$0.05}  \\
96 & {\bf 0.90$\pm$0.08} & {\bf 0.90$\pm$0.08} & {\bf 0.92$\pm$0.09} & {\bf 0.92$\pm$0.08} & {\bf 0.91$\pm$0.05} & {\bf 0.92$\pm$0.05}  \\
108 & 0.89$\pm$0.09 & 0.89$\pm$0.09 & 0.88$\pm$0.08 & 0.86$\pm$0.08 & {\bf 0.98$\pm$0.02} & {\bf 0.98$\pm$0.02}  \\
120 & 0.87$\pm$0.08 & 0.87$\pm$0.08 & 0.82$\pm$0.11 & 0.82$\pm$0.09 & {\bf 0.96$\pm$0.06} & {\bf 0.97$\pm$0.03}  \\
\hline
\end{tabular}
\end{table*}

The maximum TSS and HSS values of 0.92$\pm$0.09 and 0.92$\pm$0.08, respectively, show that the SVMs can successfully predict if a flare will be accompanied with CMEs and SEPs 96 hours prior to their occurrences. For the 36 hours forecasting window, the TSS and HSS values are 0.91$\pm$0.05 and 0.90$\pm$0.08, respectively (Table~\ref{tab:MaxTSS_HSS_SVM} and Figures~\ref{fig:TSS_SVM} and~\ref{fig:HSS_SVM}). The maximum TSS and HSS values for this class are around $\sim$0.72 on average at time delays between 48 and 72 hours, where they increase up to 0.83 for the 84 hour time delay. For time delays longer than 96 hours, the maximum TSS and HSS values are around 0.85. For the 24 hour forecast window, the TSS and HSS values are 0.85$\pm$0.14 and 0.84$\pm$0.12, respectively. The performance of the SVMs on predicting if a flare will be accompanied with CMEs and SEPs is minimum for the time delay of 12 hours (Table~\ref{tab:MaxTSS_HSS_SVM} and Figures~\ref{fig:TSS_SVM} and~\ref{fig:HSS_SVM}).

The results also show that the SVMs can predict if a CME will not accompany flares and SEPs for the time delays between 120 and 84 hours as well as 36 and 24 hours prior to their occurrences as suggested by the maximum TSS and HSS levels of above 0.90$\pm$0.10 (Table~\ref{tab:MaxTSS_HSS_SVM}). For the 12 hour time delay, the maximum TSS and HSS values are 0.82$\pm$0.11 and 0.83$\pm$0.10, respectively. The TSS and HSS values decrease down to around $\sim$0.80 for the time delays between 48 and 72 hours (Table~\ref{tab:MaxTSS_HSS_SVM}, and Figures~\ref{fig:TSS_SVM} and~\ref{fig:HSS_SVM}).

\subsubsection{Multilayer Perceptrons}
\label{sec:resultsMLP}

\begin{figure}[ht!]
\begin{center}
{\includegraphics[width=3.5in]{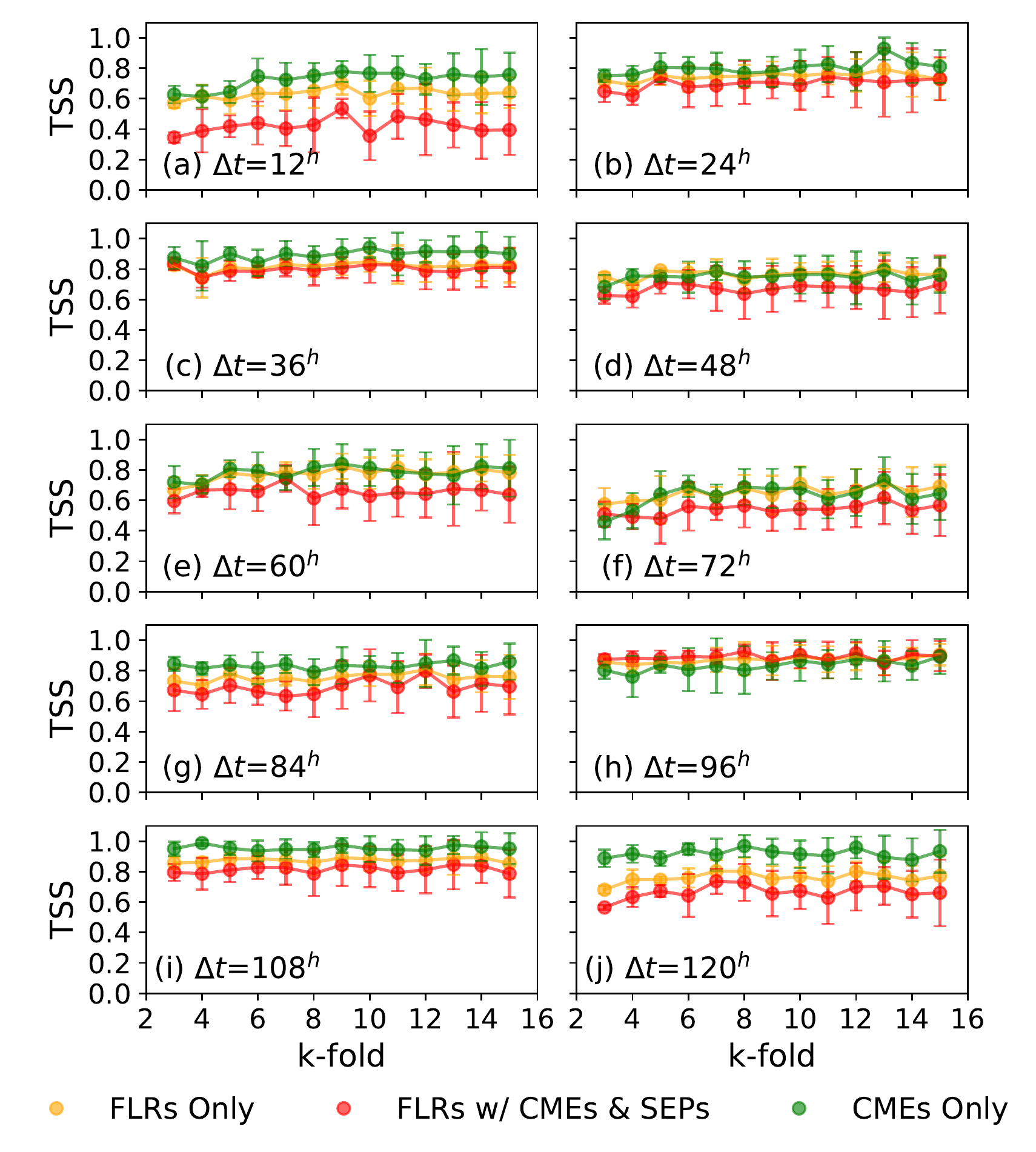}}
\caption{The TSS values calculated based on confusion matrices obtained from the MLPs as a function of $k$-folds CV and $\Delta t$ time delays for flares only (orange), flares with CMEs and SEPs (red), and CMEs only (green).} 
\label{fig:TSS_MLP}
\end{center}
\end{figure}

\begin{figure}[ht!]
\begin{center}
{\includegraphics[width=3.5in]{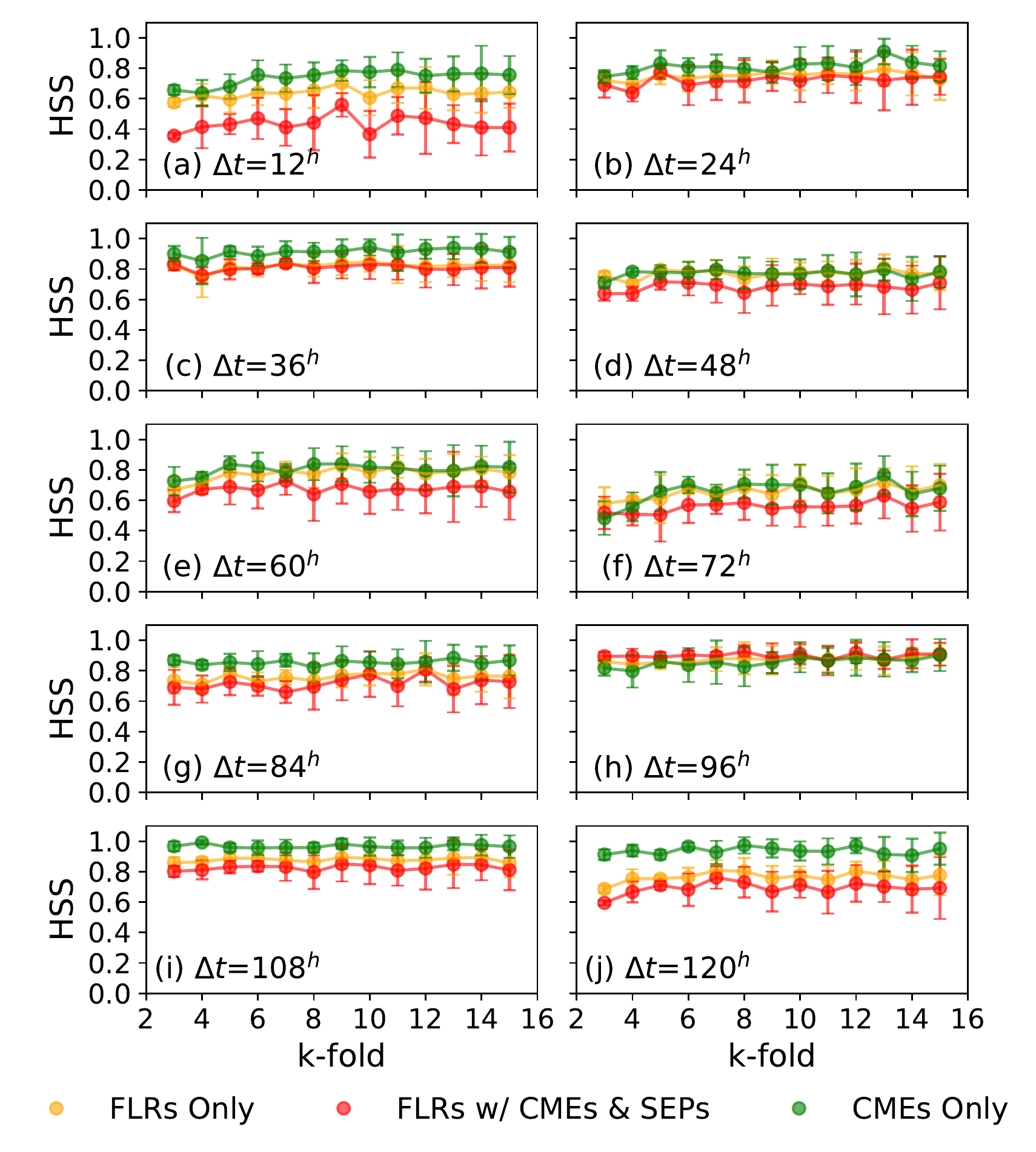}}
\caption{The HSS values calculated based on confusion matrices obtained from the MLPs as a function of $k$-folds CV and $\Delta t$ time delays for flares only (orange), flares with CMEs and SEPs (red), and CMEs only (green).} 
\label{fig:HSS_MLP}
\end{center}
\end{figure}

The TSS and HSS values are calculated based on the confusion matrices obtained from the MLPs for different $k$-fold CV. A sample for the confusion matrix for the 36 hour delay and $k=3$-fold CV show that 74 instances of the flares, 31 of the flares with CMEs and SEPs, and again 33 of the CMEs are correctly classified (Table~\ref{tab:conf_matrixMLP}). The TSS and HSS values for different $k$-fold CV suggest that the MLPs can predict that a flare will not be associated with any other event 96 hours prior to its occurrence with maximum TSS and HSS values of 0.91$\pm$0.07 (Table~\ref{tab:MaxTSS_HSS_MLP}). The TSS and HSS levels are above 0.70 for the 108 and 120 hour time delays, while they vary between 0.60 and 0.85 for the time delays between 24 and 84 hours. The TSS and HSS decrease below 0.70 for the 12 hour time delay (Figures~\ref{fig:TSS_MLP}a and~\ref{fig:HSS_MLP}a).

\begin{table}[h!]
\centering
\caption{A sample confusion matrix for one of the MLPs for the 36 hour delay, where the number of neurons is 36, the tolerance is $10^{-9}$, and $k=3$-fold CV. Indices 1, 2, and 3 denote flare, flare with CMEs and SEPs, and CMEs classes, respectively.}
\label{tab:conf_matrixMLP}
\begin{tabular}{c|ccc}

\hline
									& $O_{1}$ 	& $O_{2}$ 		& $O_{3}$  	\\
\hline
					$P_{1}$ 			& 74			&3				& 1		     \\
					$P_{2}$	 		& 6			&31				& 0			   \\
					$P_{3}$ 			& 0			&1				& 33		      \\

\hline
\end{tabular}
\end{table}

The results show that the MLPs can predict whether a flare will be accompanied with CMEs and SEPs with maximum TSS and HSS values of 0.93$\pm$0.04 for the 96 hour time delay. Further, these values are above 0.80 levels for the 108 and 120 hour time delays, while they are generally below 0.85 levels for the rest (Table~\ref{tab:MaxTSS_HSS_MLP} and Figures~\ref{fig:TSS_MLP} and~\ref{fig:HSS_MLP}).

\begin{table*}
\centering
\caption{Maximum TSS and HSS values from the MLPs for three classes at different time delays. Note that the TSS and HSS values $\ge$\ 0.90 with standard deviation $\le$\ 10\% are marked in bold face.} 
\label{tab:MaxTSS_HSS_MLP}
\begin{tabular}{ccccccccccc}
\hline
\hline
\multicolumn{1}{c}{time delay} & \multicolumn{2}{c}{flares} & \multicolumn{2}{c}{flares w/ CMEs \& SEPs} & \multicolumn{2}{c}{CMEs}   \\

\multicolumn{1}{c}{$\Delta t$ (hrs)} & TSS$_{Max}$ &HSS$_{Max}$   & TSS$_{Max}$ &HSS$_{Max}$ & TSS$_{Max}$ &HSS$_{Max}$     \\
\hline

12 & 0.70$\pm$0.07 & 0.70$\pm$0.07 & 0.54$\pm$0.06 & 0.56$\pm$0.08 & 0.78$\pm$0.07 & 0.79$\pm$0.11  \\
24 & 0.79$\pm$0.11 & 0.79$\pm$0.11 & 0.74$\pm$0.13 & 0.77$\pm$0.05 & {\bf 0.93$\pm$0.07} & {\bf 0.91$\pm$0.08}  \\
36 & 0.85$\pm$0.07 & 0.85$\pm$0.06 & 0.84$\pm$0.04 & 0.84$\pm$0.03 & {\bf 0.94$\pm$0.06} & {\bf 0.94$\pm$0.05}  \\
48 & 0.80$\pm$0.09 & 0.81$\pm$0.09 & 0.71$\pm$0.07 & 0.72$\pm$0.05 & 0.79$\pm$0.12 & 0.80$\pm$0.07  \\
60 & 0.82$\pm$0.08 & 0.83$\pm$0.08 & 0.74$\pm$0.08 & 0.73$\pm$0.09 & 0.84$\pm$0.13 & 0.84$\pm$0.12  \\
72 & 0.71$\pm$0.10 & 0.72$\pm$0.09 & 0.62$\pm$0.17 & 0.63$\pm$0.15 & 0.73$\pm$0.15 & 0.77$\pm$0.12 \\
84 & 0.81$\pm$0.11 & 0.81$\pm$0.11 & 0.80$\pm$0.11 & 0.81$\pm$0.08 & 0.87$\pm$0.09 & 0.88$\pm$0.09  \\
96 & {\bf 0.91$\pm$0.07} & {\bf 0.91$\pm$0.07} & {\bf 0.93$\pm$0.04} & {\bf 0.93$\pm$0.04} & 0.89$\pm$0.11 & 0.90$\pm$0.11  \\
108 & 0.89$\pm$0.07 & 0.89$\pm$0.07 & 0.85$\pm$0.16 & 0.85$\pm$0.11 & {\bf 0.99$\pm$0.02} & {\bf 0.99$\pm$0.01}  \\
120 & 0.80$\pm$0.07 & 0.81$\pm$0.07 & 0.74$\pm$0.08 & 0.76$\pm$0.07 & {\bf 0.97$\pm$0.07} & {\bf 0.97$\pm$0.05}  \\
\hline
\end{tabular}
\end{table*}

The maximum TSS and HSS values above 0.90$\pm$0.10 show that the MLPs can successfully predict that a CME will not be associated with flares and SEPs 120 hours, 108 hours, 36 hours, and 24 hours prior to their occurrences (Figures~\ref{fig:TSS_MLP} and~\ref{fig:HSS_MLP}). For the 108 hour delay, we obtained maximum TSS and HSS values of 0.99$\pm$0.02 and 0.99$\pm$0.01, respectively. These values gradually decrease down below 0.80 until the 48 hour delay. For the 12 hour time delay, the maximum TSS and HSS are 0.78$\pm$0.07 and 0.79$\pm$0.11, respectively (Table~\ref{tab:MaxTSS_HSS_MLP}).

\subsection{Comparison of the SVMs and MLPs}

The resulting TSS and HSS values show that the SVMs generally perform better than MLPs, having the TSS and HSS values mostly above 0.50, while the MLPs reach down to around 0.40.

The TSS and HSS values, which are calculated to predict that flares will not be associated with CMEs and SEPs, are generally higher in SVM for all $k$-fold values, while the MLP values reach down to $\sim$0.55. The two methods show overlapping high TSS and HSS values for the 96 hour delay (Figures~\ref{fig:TSS_SVM},~\ref{fig:HSS_SVM},~\ref{fig:TSS_MLP}, and~\ref{fig:HSS_MLP}).

The TSS and HSS values show that SVMs can predict that a flare will be accompanied with CMEs and SEPs in the forecast window of 96 hours (TSS=0.92$\pm$0.08, HSS=0.92$\pm$0.09), which also coincides with the highest TSS and HSS values from the MLPs (TSS=0.93$\pm$0.04, HSS=0.93$\pm$0.04). In addition to the 96 hour forecast window, the SVMs also show high TSS and HSS values for the 36 hour time delay (TSS=0.91$\pm$0.05, HSS=0.90$\pm$0.08), while the MLPs cannot reach these levels (Tables~\ref{tab:MaxTSS_HSS_SVM}, and \ref{tab:MaxTSS_HSS_MLP}). 

The SVMs can predict that a CME will not be associated with any other event with TSS and HSS values above 0.90$\pm$0.10 for the time delays continuously between 120 and 84 hours, as well as 36 and 24 hours (Table~\ref{tab:MaxTSS_HSS_SVM}). On the other hand, the MLPs give TSS and HSS values above 0.90$\pm$0.10 for the forecast windows of 120, 108, 36, and 24 hours (Table~\ref{tab:MaxTSS_HSS_MLP}).

\section{Discussion and Conclusions}
\label{sec:dis_conc}

\citet{2018SoPh..293...28F} used data from flaring and non-flaring ARs in an SVM algorithm to predict occurrences of $>$M1-class and $>$C1-class flares within a 24 hour forecasting window. For prediction of $>$M1 class flares, the authors reported TSS and HSS values of $\sim$0.72 and $\sim$0.55, respectively, while for $>$C1 class flares their TSS and HSS values decreased down to $\sim$0.57 and $\sim$0.50, respectively. To predict occurrences of flares $>$M1 class, \citet{2015ApJ...798..135B} used {\it SDO}/HMI's definitive flaring and non-flaring AR data in SVMs at time delays 48 and 24 hours, and reported TSS values of 0.82 and 0.76, respectively. In addition, again based on the {\it SDO}/HMI's definitive AR data that produce flares and flares with associated CMEs in SVMs, \citet{2016ApJ...821..127B} calculated TSS=$\sim0.80\pm0.20$ to predict whether a flare will be associated with CMEs in a 24 hour forecast window. On the other hand, \citet{2015ApJ...812...51B} used the {\it SOHO}/MDI data of flaring and non-flaring ARs in an SVM regressor and calculated TSS=0.55 and HSS=0.46 in their size regressions for $>$C1 class flares, which is used to predict the size of a flare. To predict flare classes A, B, C, M, and X, \citet{2010RAA....10..785Y} used SVMs and achieved maximum TSS and HSS values of 0.63 and 0.64 for predictions of A and B class flares, 0.09 and 0.11 for only C class flares, 0.05 and 0.06 for only M class flares, and 0.14 and 0.18 for only X class flares, respectively. We recovered the TSS and HSS values based on the confusion matrices given in their work \citep[Figures 3b, 4b, 5b, and 6b in][]{2010RAA....10..785Y}.

Using SVMs for the 24 hour forecast window, we reached maximum TSS and HSS values of 0.85$\pm$0.10 to predict whether a flare will be unassociated with any other event. The maximum TSS and HSS values to predict whether a flare will be accompanied by CMEs and SEPs are 0.85$\pm$0.14 and 0.84$\pm$0.12, respectively. As for predicting whether an AR will produce only CMEs, our maximum TSS and HSS values are 0.91$\pm$0.10 (Table~\ref{tab:MaxTSS_HSS_SVM}). Additionally, our maximum TSS and HSS values from the SVMs for the 48 hour prediction window that a flare will not be associated with any other event is 0.77$\pm$0.08. For predicting whether a flare will be associated with CMEs and SEPs, our TSS=0.80$\pm$0.10 and HSS=0.80$\pm$0.09, while for forecasting that a CME will not be associated with any other event, the maximum TSS and HSS values are 0.79$\pm$0.19 and 0.80$\pm$0.09, respectively. Among the time delays we used in our study, which range from 120 to 12 hours, the forecast windows of 96 and 36 hours consistently provided maximum TSS and HSS values above 0.90 with standard deviations smaller than 10\% (Table~\ref{tab:MaxTSS_HSS_SVM}). These results are better than those found by the previous studies. However, it must be noted that the scope of this study is to distinguish the three classes of solar eruptions observed on the Sun, therefore non-flaring ARs are not included. Also, the choice of the parameters used in this study differs from most of the previous studies.

Using MLPs to predict the occurrences of $>$M1 and $>$C1 class flares within a 24 hour forecast window, separately, \citet{2018SoPh..293...28F} obtained maximum TSS and HSS values of $\sim$0.73 and $\sim$0.55 for $>$M1 class flares, and $\sim$0.57 and $\sim$0.55 for $>$C1 class flares, respectively \citep[Figures 3 and 5 in][]{2018SoPh..293...28F}. Using a cascade correlation neural network algorithm to predict the occurrences of $>$C1 class flares within 24 and 48 hour forecast windows, \citet{2013SoPh..283..157A} calculated maximum TSS and HSS as 0.45 and 0.54, respectively. \citet{2009SoPh..255...91Y}, on the other hand, used a learning vector quantisation neural networks algorithm to predict flare occurrences within a forecast window of 48 hours and reached maximum TSS=0.67, which we recovered using $TSS=TP_{rate}-(1-TN_{rate})$ relationship.

Our maximum TSS and HSS values obtained from the MLPs for the 24 hour forecast window are 0.79$\pm$0.11 for forecasting that a flare will not be associated with any other event. To predict whether a flare will be associated with CMEs and SEPs, we obtained TSS=0.74$\pm$0.13 and HSS=0.77$\pm$0.05, while these values are 0.93$\pm$0.07 and 0.91$\pm$0.08 for forecasting that a CME will not be associated with any other event (Table~\ref{tab:MaxTSS_HSS_MLP}). As for the forecast window of 48 hours, we obtained TSS=0.80$\pm$0.09 and HSS=0.81$\pm$0.09 for predictions of flare only events. The MLPs gave maximum TSS and HSS values of 0.71$\pm$0.07 and 0.72$\pm$0.05 for predictions whether a flare will be accompanied with CMEs and SEPs, while these numbers are 0.79$\pm$0.12 and 0.80$\pm$0.07 for predictions of CME only events. Although our maximum TSS and HSS values are higher than those found in the previous studies, we must again note that we do not include data from non-flaring ARs, since our main focus here is to distinguish the three classes of solar eruptions observed on the Sun.

In our study, we used SVMs and MLPs to predict whether an AR, which is known to originate solar eruptions, will produce only flares, flares with associated CMEs and SEPs, or only CMEs for time delays extending from 12 hours to 120 hours. To achieve this objective, we used data provided by the SHARPs, {\it GOES} and DONKI databases. The size of the data used in this study is limited due to the mission duration of {\it SDO}/HMI and also because of the quietness of the current solar cycle 24, which leads to fewer flares, CMEs and SEPs. To overcome this difficulty, we employed the stratified $k$-fold CV method, which ensures that each subset of data is used to train and test the ML algorithms. However, this method assumes that the underlying data follows the same distribution throughout the study period. It is shown that the number of occurrences of flares, CMEs and SEPs, although sporadic, follow a trend in-phase with the 11-year solar cycle \citep{2010ApJ...717..683A,2011LRSP....8....1C,2015LRSP...12....4H}. To decrease this bias in our calculations, we shuffled the data before performing $k$-fold CV method. However, to remove this bias substantially, to avoid overfitting, and to train and validate the ML algorithms further, longer data sets are needed.

To optimise the hyper-parameters used in our SVM and MLP algorithms, we employed an embedded grid search for each iteration of $k$-fold CV values ranging from 3 to 15. Our results show that we can achieve TSS and HSS values greater than 0.90 with standard deviations smaller than 10\%, which shows both our SVM and MLP are good classifiers, though the former is slightly better than the latter. Our results also show that we can predict that flares will not be accompanied with any associated event earliest 96 hours before they happen with maximum TSS and HSS values of 0.90$\pm$0.08 for the SVMs and 0.91$\pm$0.07 for the MLPs. The earliest forecast that a flare will be associated with CMEs and SEPs can be made at 96 hour time delay with TSS=0.92$\pm$0.09 and HSS=0.92$\pm$0.08 for the SVMs and TSS=HSS=0.93$\pm$0.04 for the MLPs. We also showed that we can predict if an AR will produce only CMEs unassociated with any other events 108 hours before they occur with maximum TSS and HSS values of 0.98$\pm$0.02 for the SVMs and 0.99$\pm$0.02 for the MLPs. Our results indicate that the discriminative potential of the physical features of ARs in SHARPs data is very high.

We also calculated the Fisher-scores of the 18 physical parameters for each time delay, which are not shown in this study, as we do not include them as feature selection criteria. However, we need to note that the results show that the variation in the Fisher-scores of the physical parameters of the ARs provide insights into why the predictive powers of the SVMs and MLPs change with different time delays. Additionally, the calculated Fisher scores indicate that the physical features are highly complementary across the different time delays, meaning that all features are relevant, but not necessarily at the same time delay.

In conclusion, our results show that the SVMs are slightly better than the MLPs. However, a more extensive future work is necessary for SVM classifiers, where ARs that do not produce any eruptive phenomena are planned to be included in the analyses. This will, however, introduce imbalance problems because the number of ARs that do not produce an eruptive event is overwhelmingly high compared to those produce eruptions \citep{2015ApJ...798..135B}. It is therefore important to find the optimum values for the class weight ratios to sort out this problem. Additionally, we plan to investigate the optimum number of features by combining the predictions from the individual time delays, to search a larger grid for the C and $\gamma$ values, and to employ different kernel functions and grid search their related parameters (such as polynomial functions with different degrees). Furthermore, another direction we plan to exploit is using deep neural networks with temporal evolutionary algorithms on the time-series aspect of the available SHARPs data.

\begin{acknowledgements}
Funding for the Stellar Astrophysics Centre is provided by the Danish National Research Foundation (Grant agreement No.: DNRF106). The project has been supported by the Villum Foundation.
\end{acknowledgements}

\clearpage

\clearpage

\end{document}